\documentclass[prc,twocolumn,preprintnumbers,amsmath,amssymb,superscriptaddress,floatfix,nofootinbib]{revtex4}
\usepackage{graphicx}
\usepackage{epsfig}
\usepackage{amsmath}
\newcommand{\slashed}[1]{#1\!\!\!/}

\newcommand{\tabincell}[2]{\begin{tabular}{@{}#1@{}}#2\end{tabular}}

\def\prl#1{Phys.\ Rev.\ Lett.\ {\bf #1}}

\def\prc#1{Phys.\ Rev.\ C\ {\bf #1}}

\def\etal{{\em et al.}}

  \def\CL{{\cal L}}
\def\CM{{\cal M}}

\def\be{\begin{equation}}
\def\ee{\end{equation}}
\def\Be{\begin{eqnarray}}
\def\Ee{\end{eqnarray}}
\def\ba{\begin{array}}
\def\ea{\end{array}}

\begin{document}
\title{$\Delta(1232)$ production in the $\gamma p \to \phi\pi p$ reaction}

\author{Ya-Ke Chen}
\author{Yi Pan}
\author{Bo-Chao Liu}
\email{liubc@xjtu.edu.cn}

\address{Ministry of Education Key Laboratory for Nonequilibrium Synthesis
and Modulation of Condensed Matter, Shaanxi Province Key Laboratory of Quantum Information and
Quantum Optoelectronic Devices, School of Physics, Xi’an Jiaotong University, Xi’an 710049, China}

\begin{abstract}
In this work, the production of baryon resonances in the $\gamma p \to \phi \pi p$ reaction near threshold is investigated within an effective Lagrangian approach.  incorporates intermediate resonances, including $\Delta(1232)$, $N(1440)$, $N(1520)$, and $N(1535)$, alongside non-resonant background contributions. The results indicate that the reaction is dominated by the $\Delta(1232)$ excitation near threshold. A key focus of this study is the exploration of the $a_0(980)$ meson exchange in the $\Delta(1232)$ production, as this reaction offers a unique opportunity to probe the poorly known $\Delta(1232)N a_0$ coupling. We demonstrate that the parity asymmetry ($P_\sigma$), accessible through the $\phi$-meson's spin density matrix elements, serves as a suitable experimental observable to identify the contribution from the $a_0(980)$ exchange.

\end{abstract}
\maketitle

\section{Introduction}
Investigating baryon resonances plays a crucial role in hadronic physics, shedding light on the non-perturbative dynamics of Quantum Chromodynamics (QCD). Due to their extremely short lifetimes, resonances such as the \(\Delta(1232)\) cannot be observed directly. Instead, their existence and properties are inferred from analyses of intermediate states in scattering or decay processes. The \(\Delta(1232)\) resonance, characterized by its prominent peak in both \(\pi N\) and \(\gamma N\) scattering reactions, serves as a particularly important archetype for these studies. This is primarily due to its relatively low mass and strong coupling to the \(\pi N\) channel \cite{pdg2024}.
However, the low mass of the \(\Delta(1232)\) also presents significant challenges when investigating its coupling to channels involving heavier particles. A key limitation arises because the \(\Delta(1232)\) mass lies below the kinematic thresholds required to produce the heavier states. Consequently, decay channels that would probe such couplings: such as \(\Delta \to N\rho\), \(\Delta \to Na_0(980)\), and \(\Delta \to K\Sigma\), are kinematically forbidden. This physical constraint precludes direct experimental investigation of these specific interactions through the \(\Delta(1232)\) decays.
As a result, the couplings of the \(\Delta(1232)\) resonance to heavier particle systems, such as \(N\rho\)\cite{Xie:2008ts}, \(Na_0(980)\), and \(K\Sigma\), remain poorly constrained or entirely unknown. These significant gaps in our knowledge considerably hamper efforts to achieve a comprehensive understanding of the \(\Delta(1232)\)'s dynamical properties and its possible role within the broader landscape of hadronic interactions.

In recent years, photoproduction reactions have emerged as a powerful tool for investigating the properties of baryon resonances~\cite{Thiel:2022xtb,Mokeev:2022xfo,Ireland:2019uwn,Burkert2018,Anisovich2017,Mattione2017,Kamano2013,Gothe2016}. 
While most studies have focused on single-meson production processes, extending investigations to multi-meson final states---for example, the $N\pi\pi$, $NK\bar{K}$, $N\eta\pi$ channels etc.~\cite{CBELSATAPS:2015kka,A2:2015pgk,A2:2016vzp,CBELSATAPS:2014wvh,A2:2018vbv,CLAS:2018azo,CLAS:2018drk}---offers unique opportunities to probe resonance characteristics. Furthermore, it was demonstrated in Ref.~\cite{lv} that the reaction $\gamma p \to \phi K^+\Lambda$, owing to its specific production dynamics, is particularly well-suited for studying the $N(1535)K\Lambda$ coupling\cite{oset2002,Liu2006}. Probing this specific coupling through single-meson production processes presents significant challenges. 
This example, along with other works~\cite{Liu2017,Liu2019,Liu2020,Zhao:2019syt,Dai:2025hvo}, demonstrate that the dynamics of multi-meson final states provide a unique avenue to study resonance couplings.

In this paper, we present the results of a study of the $\gamma p\to \pi \phi p$
 reaction near threshold using an effective Lagrangian approach. We
consider the contributions from the $\Delta(1232)$, $N(1440)$, $N(1520)$ and $N(1535)$ resonances as intermediate states, which subsequently
decay into the $\pi N$ in the final state. Contributions from resonances in the $\phi\pi$ and $\phi p$ channels are ignored in
this work, because no significant resonance signals have been found in these two channels in the energy region under study\cite{pdg2024}. This observation suggests that the present reaction is well suited for studying the excitation mechanisms of the aforementioned baryon resonances in the
$\gamma p \to \phi \Delta$ or $\phi N^*$ processes since their decays into the $\pi N$ channel are well known. Such studies are not only
important for understanding the reaction mechanism itself but also
valuable for probing the coupling of baryon resonances with
exchanged particles. For the $t$-channel of the $\gamma p \to \phi \Delta$ or $\phi N^*$ subprocesses, vector meson exchange is forbidden due to the conservation of C-parity. Since the nucleon and $\Delta$ resonances considered in this work
have relatively large decay branch ratios to the $N\pi$
channel, it is natural to expect that the $\pi$-meson
exchange plays an important role in the excitation of the $\Delta(1232)$ and $N^*$s in
this reaction. Furthermore, for the $N(1535)$ production, $\eta$-meson exchange could also be important. Other possible contributions, such as the exchange of scalar mesons($\sigma$, $f_0(980)$ and
$a_0(980)$) and axial vector mesons($a_1(1260)$,$b_1(1235)$), may also exist. These
contributions were usually ignored in previous studies due to their
relatively large mass or their poorly known couplings with the $\Delta(1232)$ or
$N^*$s. These assumptions require experimental verification.

In this work, we try to investigate the role of
$a_0(980)$, hereafter denoted as $a_0$, in
the excitation of the $\Delta(1232)$ in this reaction. To this end, we note that since the isospin of $\Delta(1232)$ is 3/2, isoscalar meson exchanges are forbidden in this process. Meanwhile, the PDG book\cite{pdg2024} shows that axial vector mesons such as $a_1(1260)$ and $b_1(1235)$ do not have a significant decay branch ratio to the $\phi\gamma$ channel, which means their contributions are suppressed to some extent. Considering this and their relatively large masses, we neglect their contributions. Furthermore, we will focus on the near-threshold region, where the production of $\Delta(1232)$ is expected to be dominant. To estimate the
$a_0$ exchange contribution, knowledge of the $\phi a_0\gamma$
and $\Delta Na_0$ couplings is essential. The $\phi a_0\gamma$
coupling can be extracted from the radiative decay width of $\phi$ to $\gamma
a_0$\cite{Kalashnikova2005}. However, the coupling of $\Delta(1232)$
with $Na_0$ has been rarely studied in previous studies. Therefore, if the $a_0$ exchange contribution can be identified experimentally, it offers an opportunity to study the $\Delta(1232)N a_0$ coupling. To do this, a crucial step is to separate the $a_0$ exchange contribution from that of the $\pi$ exchange. This problem can be solved by following the method suggested in previous works\cite{Tannoudji,Schilling}, where it was shown that spin observables are very useful for this purpose. The spin density matrix elements(SDMEs) of the $\phi$ meson can be
extracted from its decay angular distribution. By analyzing the
SDMEs of the $\phi$ meson, it is possible to obtain the information
about the exchanged particles. This method was already used in
Refs.\cite{Oh2006,Oh2012} to identify the role of the scalar $\kappa$
meson exchange. In this work, we will show that the SDMEs can also
provide useful information about the exchanged meson in the $\Delta(1232)$
production process.

The rest of this paper is organized as follows. In Sec.II, we
present the model and the theoretical formalism used in the calculations.
Numerical results and discussions are presented in Sec. III,
followed by a summary in the last section.
\begin{figure*}[htbp]
\centering
\includegraphics[scale=0.45]{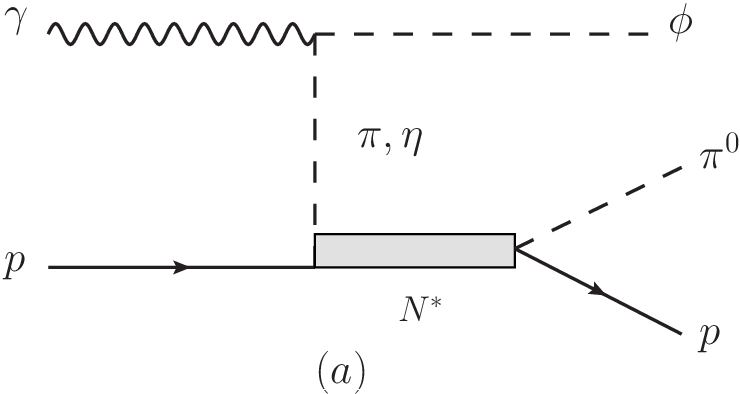}
\includegraphics[scale=0.45]{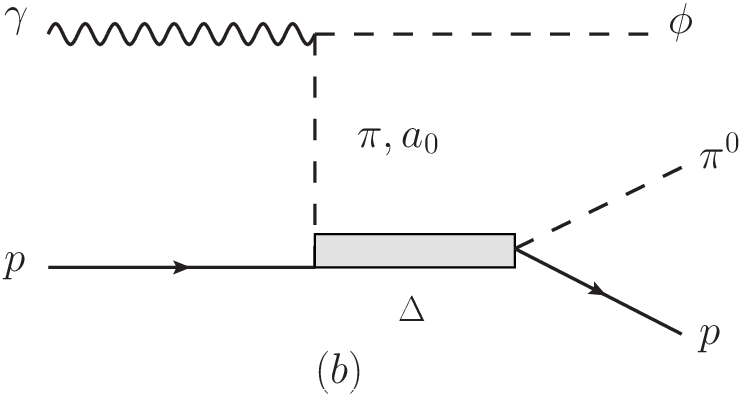}
\includegraphics[scale=0.45]{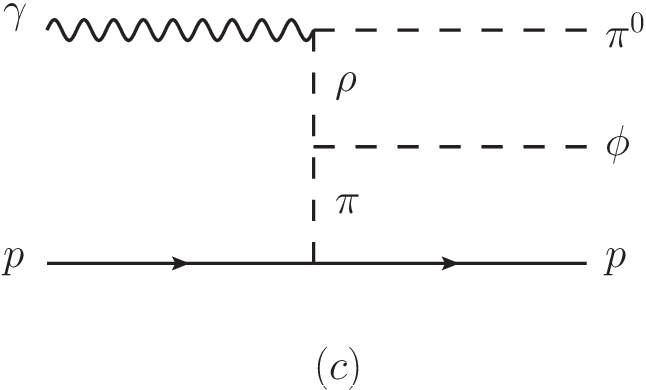}
\caption{Feynman diagrams for the ${\gamma}p \to {\phi}{\pi}p$
reaction.} \label{feynfig}
\end{figure*}

\section{the model}
The Feynman diagrams for the $\gamma p \to \phi \pi p$ reaction in our model are depicted in Fig.\ref{feynfig}. Since we focus on the near-threshold region, resonance states such as $\Delta(1232)$, $N(1440)$, $N(1520)$, and $N(1535)$ in the intermediate states are included. As mentioned in the \textit{Introduction}, we consider the $\pi$ and $\eta$ exchanges for nucleon resonance excitations, and the $\pi$ and $a_0$ exchanges for the excitation of the $\Delta(1232)$ resonance, as other meson exchanges are either forbidden or considered unimportant. To calculate the Feynman diagrams shown in Fig.\ref{feynfig}, the following Lagrangian densities are needed\cite{lv,Lee2017,Zou2003}: \Be
\mathcal{L}_{{\pi}NN} &=& - \frac{{{g_{\pi NN}}}}{{2{m_N}}}\bar N {\gamma _5}{\gamma _\mu }{\partial ^\mu }{\pi}N,\\
\CL_{\gamma {\pi}\phi } &=& {e \over {{m_\phi }}}{g_{\phi \gamma \pi
}}{\varepsilon ^{\mu \nu \alpha \beta }}{\partial _\mu }{\phi
_\nu }{\partial _\alpha }{A_\beta }{\pi},\\
\CL_{\gamma \eta \phi} &=& {e \over {{m_\phi }}}{g_{\phi \gamma \eta
}}{\varepsilon ^{\mu \nu \alpha \beta }}{\partial _\mu }{\phi _\nu
}{\partial _\alpha }{A_\beta }\eta,\\
\CL_{\rho_0 \pi \gamma} &=& \frac{e}{{{m_\rho }}}{g_{\rho \pi
\gamma}}{\varepsilon ^{\mu \nu \alpha \beta }}{\partial _\mu
}{\rho_\nu }{\partial _\alpha }A_\beta \pi,\\
\CL_{\phi \rho_0 \pi} &=& \frac{g_{\phi \rho
\pi}}{{{m_\phi }}}{\varepsilon ^{\mu \nu \alpha \beta }}{\partial _\mu
}{\phi_\nu }{\partial _\alpha }\rho_\beta \pi,\\
\CL_{\phi a_0 \gamma}&=&\frac{e}{m_\phi}g_{\phi a_0 \gamma}{\partial
^\alpha}{\phi ^\beta}({\partial _\alpha}A_\beta-{\partial
_\beta}A_\alpha)a_0,\\
\CL_{\Phi N{N_{1535}^*}} &=& {g_{\Phi N{N_{1535}^*}}}{\bar N^*}MN + h.c.,\\
\CL_{\pi NN_{1440}^*} &=&  - \frac{g_{\pi NN_{1440}^*}}{{m_N} +
{m_{N^*}}}{\bar{N}^*}{\gamma _5}{\gamma _\mu }{\partial ^\mu }MN
+ h.c.,\\
\CL_{\pi NN_{1520}^*} &=&  -
\frac{{{g_{MNN_{1520}^*}}}}{{{m_\pi}}}{\bar{N}}_\mu ^*{\partial ^\mu
}M\gamma_5N + h.c. ,\\
\mathcal{L}_{\Delta N a_0} &=& \frac{g_{\Delta Na_0} }{m_{a_0}}\bar\Delta^{\mu} (\vec\tau \cdot \partial_\mu \vec{a}_0) \gamma_5 N + h.c.\\
\mathcal{L}_{\Delta N\pi} &=& \frac{g_{\Delta N\pi } }{m_{\pi}} \bar\Delta^{\mu}(\vec\tau \cdot \partial_\mu \vec{\pi}) N + h.c.\Ee where $\phi_\mu$ is the $\phi$ meson field, $\Phi$ stands for the $\pi$ or $\eta$ fields, and $A_\mu$ is the photon field. The coupling constant $g_{\pi NN}$ is taken from Refs.\cite{Xie2008,Xie2013} with a value of $g_{\pi NN}=13.45$. The elementary charge $e$ is taken as $\sqrt{4\pi/137}$. Other coupling constants can be calculated through the partial decay width using the following formulae: \Be
\Gamma [V \to \Phi \gamma] &=& \frac{{{e^2}g_{V \Phi \gamma}^2}}{12\pi
}\frac{{\left| {\vec{p}}
\right|}^3}{m_V ^2},\\
\Gamma [\phi  \to \rho_0 \pi] &=& \frac{{{e^2}g_{\phi \rho
\pi}^2}}{12\pi }\frac{{\left| {\vec{p}}
\right|}^3}{m_\phi ^2},\\
\Gamma [\phi \to a_0 \gamma]&=&\frac{{e^2}g_{\phi a_0
\gamma}^2}{{12\pi }}\frac{{{{\left| {{\vec{p}}}
\right|}^3}}}{{m_\phi ^2}}\label{f0phigamma},\\
\Gamma [{N(1535)} \to \Phi N] &=& \frac{\kappa
{g_{\Phi N{N^*}}^2}}{{4\pi }}\frac{{({E_N} +
{m_N})}}{{{m_{{N^*}}}}}\left| {{\vec{p}}} \right|,\\
\Gamma [{N(1440)} \to \pi N] &=& \frac{3
{g_{PN{N^*}}^2}}{{4\pi }}\frac{({E_N} -
{m_N})}{m_{N^*}}\left| {{\vec{p}}} \right|,\\
\Gamma [{N(1520)} \to \pi N] &=& \frac{
{3g_{\pi N{N^*}}^2}}{{4\pi }}\frac{{({E_N} -
{m_N})}}{{m_{N^*}}m_{\pi}^2}{\left| {{\vec{p}}} \right|}^3,\\
\Gamma [{\Delta(1232)} \to \pi N] &=& \frac{
{3g_{\pi N\Delta}^2}}{{4\pi }}\frac{{({E_N} +
{m_N})}}{{m_{\Delta}}m_{\pi}^2}{\left| {{\vec{p}}} \right|}^3, \Ee where
$|\vec{p}|$ denotes the magnitude of the three-momentum of the final particles  in the
center-of-mass(CM) frame. P stands for the $\pi$ or $\eta$ meson, and V represents the $\rho$ or $\phi$
meson. $\kappa$ is an isospin factor which equals 1 for the
$\eta$ meson and 3 for the $\pi$ meson. Using the central values from the PDG\cite{pdg2024} for the partial decay widths, the coupling
constants are calculated and listed in Tab.\ref{cc}.

\begin{table}[htbp]
\centering \caption{Coupling constants used in this work. The
experimental branch ratios are taken from PDG\cite{pdg2024}.}
\label{cc}
\begin{tabular}{ccccc}
\hline 
\hline 
State&\tabincell{c}{Width\\(Mev)}&\tabincell{c}{Decay\\channel}&\tabincell{c}{Adopt\\branching
ratio}&$g^2/4\pi$\\
\hline  
$\rho_0$&$147.8$&$\pi \gamma$&$4.7 \times 10^{-4}$&$2.6\times 10^{-2}$\\
$\phi$&$4.25$&$\pi \gamma$&$1.32 \times 10^{-3}$&$1.60 \times 10^{-3}$\\
&&$\eta \gamma$&$1.3 \times 10^{-2}$&$3.97 \times 10^{-2}$\\
&&$\rho \pi$&$0.15$&$0.326$\\
&&$a_0 \gamma$&$7.6 \times 10^{-5}$&$0.20$\\
$\Delta(1232)$&$117$&$N \pi$&$0.994$&$0.12$\\
$N(1440)$&$350$&$N \pi$&$0.65$&$3.37$\\
$N(1520)$&$110$&$N \pi$&$0.60$&$0.19$\\
$N(1535)$&$150$&$N \pi$&$0.42$&$3.43 \times 10^{-2}$\\
&&$N \eta$&$0.42$&$0.28$\\
\hline  
\hline  
\end{tabular}

\end{table}

Since hadrons are not point-like particles, it is necessary to introduce form factors at the interaction vertices. For the baryon resonance exchange diagrams, we use the following form factor, as in Refs. \cite{Shklyar2005,Xie2012}:
\begin{equation}
{F_B}(q_{ex},{m_{ex}}) = \frac{{\Lambda _B^4}}{{\Lambda _B^4 +
{{(q_{ex}^2 - m_{ex}^2)}^2}}}.
\end{equation}

For the $\pi$, $\eta$ and $a_0$ meson exchanges, we
adopt the form factor from Ref.\cite{Oh2008}:
\begin{equation}
{F_M}(q_{ex},{m_{ex}}) = {(\frac{{\Lambda _M^2 -
m_{ex}^2}}{{\Lambda _M^2 - q_{ex}^2}})^2}.
\end{equation}
For the $\rho$ meson, we use the following form factor from Ref.\cite{Oh2011}:
\begin{equation}
{F_V}(q_{ex}) = {(\frac{{\Lambda _V^2}}{{\Lambda _V^2 -
q_{ex}^2}})^2}
\end{equation}
Here, $q_{\text{ex}}$ and $m_{\text{ex}}$ are the four-momentum and mass of the exchanged particle, respectively. For the cutoff parameters, we take $\Lambda_{\pi} = \Lambda_{\eta} = 1.3$ GeV and $\Lambda_{\rho} = 1.2$ GeV for meson exchanges \cite{Liu2017,Xie2008}, and $\Lambda_B = 2.0$ GeV \cite{lv} for baryon exchanges.

The propagators for the exchanged particles are given as follows. For spin-0 mesons($\pi$ and $\eta$):
\begin{equation}
{G_0}(q) = \frac{i}{{{q^2} - m^2}}.
\end{equation}
For spin-1 $\rho$ meson:
\begin{equation}
{G_1^{\mu\nu}}(q) = -\frac{i(g^{\mu \nu}- {q^\mu
q^\nu}/{q^2})}{{q^2} - m^2}.
\end{equation}
For spin-1/2 baryon resonances:
\begin{equation}
G_{\frac{1}{2}}(q) = \frac{i(\slashed{q} + m)}{{q^2} - {m^2} +
im\Gamma }.
\end{equation}
For the spin-3/2 baryon resonances:
\begin{equation}
G_{\frac{3}{2}}^{\mu \nu }(q) = \frac{{i(\slashed{q} + m){P^{\mu \nu
}}(q)}}{{{q^2} - {m^2} + im\Gamma }},
\end{equation}
with the $P^{\mu \nu }$ given by
\begin{equation}
{P^{\mu \nu }}(q) =  - {g^{\mu \nu }} + \frac{1}{3}{\gamma ^\mu
}{\gamma ^\nu } + \frac{1}{{3m}}({\gamma ^\mu }{q^\nu } - {\gamma
^\nu }{q^\mu }) + \frac{2}{{3{m^2}}}{q^\mu }{q^\nu }.
\end{equation} Here q, m, and $\Gamma$ are the four-momentum, mass and
width of the exchanged particle, respectively.

Using the above effective Lagrangian densities and propagators, we obtain the scattering amplitudes of the Feynman diagrams shown in Fig.\ref{feynfig}. Here the momenta of individual particles in the reaction $\gamma(p_1) + p(p_2) \to \phi(p_3) + \pi(p_4) + p(p_5)$ are indicated in parentheses. The corresponding amplitudes involving nucleon resonances(Fig.\ref{feynfig}a) are
\begin{equation}
\begin{aligned}
\CM_{N^*(1535)} =& \frac{e{g_{\phi \gamma P
}}{g_{{N^*}NP}}{g_{{N^*}N\pi}}}{m_\phi}{\bar{u}}({p_5},{s_5})G_{\frac{1}{2}}(q_{N^*})\\
&\times  u({p_2},{s_2}) {\varepsilon ^{\mu \nu \alpha \beta
}}{p_{3\mu }}\varepsilon _\nu ^*({p_3},{s_3}){p_{1\alpha
}}\varepsilon _\beta ({p_1},{s_1})\\
&\times \frac{{F_B}(q_{N^*},{m_{N^*}}){F_M}(q_{P},{m_{P}})}{{{{({p_3} - {p_1})}^2} -
m_P^2}},
\end{aligned}
\end{equation}

\begin{equation}
\begin{aligned}
\CM_{N^*(1440)} =& \frac{{e{g_{\phi \gamma \pi }}{g_{{N^*}N\pi
}^2}}}{{{m_\phi}(m_p+m_{N^*})}}{\bar{u}}({p_5},{s_5}){\gamma
_5} {\slashed{p}_4}G_{\frac{1}{2}}(q_{N^*})\\
\times&{\gamma_5}({\slashed{p}_3} - {\slashed{p}_1})u({p_2},{s_2})
\frac{{F_B}(q_{N^*},{m_{N^*}}){F_M}(q_{\pi},{m_{\pi}})}{{{({p_3} - {p_1})}^2} - m_\pi ^2}\\
&\times{\varepsilon ^{\mu \nu \alpha \beta }}{P_{3\mu }}\varepsilon
_\nu ^*({p_3},{s_3}){p_{1\alpha }}\varepsilon _\beta ({p_1},{s_1}),
\end{aligned}
\end{equation}

\begin{equation}
\begin{aligned}
{\CM_{N^*(1520)}} =& \frac{{e{g_{\phi \pi\gamma
}}{g_{N^* N\pi}^2}}}{{{m_\phi }{m_\pi^2
}}}{\bar{u}}({p_5},{s_5}){p_{4\mu
}}\gamma_5 G_{\frac{3}{2}}^{\mu \nu }(q_{N^*})\gamma_5\\
\times &{({p_1} - {p_3})_\nu }u({p_2},{s_2})
\frac{{F_B}(q_{N^*},{m_{N^*}}){F_M}(q_{P},{m_{P}})}{{{{({p_3} - {p_1})}^2} - m_P^2}}\\
&\times{\varepsilon ^{\mu \nu \alpha \beta }}{p_{3\mu }}\varepsilon
_\nu ^*({p_3},{s_3}){p_{1\alpha }}\varepsilon _\beta ({p_1},{s_1}),
\end{aligned}
\end{equation}
where P denotes the exchanged $\pi$ or $\eta$. The amplitudes for the $\Delta(1232)$ production process(Fig.\ref{feynfig}b) are 
\begin{equation}
\begin{aligned}
{\CM_{{\Delta}}^\pi} =& \frac{{2e{g_{\phi \pi\gamma
}}{g_{\Delta N\pi}^2}}}{{{m_\phi }{m_\pi^2
}}}{\bar{u}}({p_5},{s_5}){p_{4\mu
}}G_{\frac{3}{2}}^{\mu \nu }(q_{\Delta})\\
\times &{({p_1} - {p_3})_\nu }u({p_2},{s_2})
\frac{{F_B}(q_{\Delta},{m_{\Delta}}){F_M}(q_{
\pi},{m_{\pi}})}{{{{({p_3} - {p_1})}^2} - m_\pi^2}}\\
&\times{\varepsilon ^{\mu \nu \alpha \beta }}{p_{3\mu }}\varepsilon
_\nu ^*({p_3},{s_3}){p_{1\alpha }}\varepsilon _\beta ({p_1},{s_1}),
\end{aligned}
\end{equation}
\begin{equation}
\begin{aligned}
\CM_\Delta^{a_0} =& \frac{-e{g_{\phi \gamma a_0
}}{g_{\Delta Na_0}}{g_{\Delta N\pi}}}{m_\phi m_{a_0}m_\pi}{\bar{u}}({p_5},{s_5})p_{4\mu} G^{\mu\nu}_{\frac{3}{2}}(q_\Delta)\\
&\times  \gamma_5 u({p_2},{s_2})(p_1-p_3)_\nu [(p_3 \cdot p_1)(\varepsilon_3^* \cdot \varepsilon_1)\\
&-(p_3 \cdot \varepsilon_1)(p_1 \cdot \varepsilon_3^*)]\times \frac{{F_B}(q_{\Delta},{m_{\Delta}}){F_M}(q_{a_0},{m_{a_0}})}{{{({p_3} - {p_1})}^2} -
m_{a_0}^2}.
\end{aligned}
\end{equation}
The
corresponding amplitude for Fig.\ref{feynfig}c can be written as
\begin{equation}
\begin{aligned}
\CM_t =& -\frac{{{ie}{g_{\phi \rho \pi }}{g_{\pi \rho \gamma
}}{g_{\pi NN}}}}{{2{m_p}{m_\phi }{m_\rho
}}}{\bar{u}}({p_5},{s_5}){\gamma _5}({\slashed{p}_2} -
{\slashed{p}_5})u({p_2},{s_2})\\
&\times \frac{{F_V}(q_{\rho}){F_M}(q_{\pi},{m_{\pi}})}{{{{({p_5} -
{p_2})}^2} - m_\pi ^2}}{\varepsilon ^{\mu \nu \alpha \beta
}}{p_{3\mu }}\varepsilon _\nu
^*({p_3},{s_3}){({p_1} - {p_4})_\alpha }\\
&\times {G_{1,{\beta \delta}}}(p_4-p_1)
{\varepsilon ^{\rho\sigma\gamma\delta}}{p_{1\rho}}{\varepsilon
_\sigma}({p_1},{s_1}){({p_1} - {p_4})_\gamma}.
\end{aligned}
\end{equation}

With these individual amplitudes, the differential cross section for this reaction is calculated by integrating over the three-body final state phase space:
\begin{equation}
d\sigma = \frac{1}{4} \frac{1}{4(p_1 \cdot p_2)} \sum_{\text{spins}} |\mathcal{M}|^2 d\Phi_3,
\end{equation}
where the summation is over the final state spins and an average is taken over the initial state spins of the photon and proton. The full scattering amplitude $\CM$ is a coherent sum of the individual amplitudes. The Lorentz-invariant phase space element 
is given by
\Be
d\Phi_3 = (2\pi)^4 \delta^{(4)} (p_1 + p_2 - p_3 - p_4 - p_5)  \nonumber\\\times\frac{d^3 \mathbf{p}_3}{(2\pi)^3 2E_3}\frac{d^3 \mathbf{p}_4}{(2\pi)^3 2E_4} \frac{d^3 \mathbf{p}_5}{(2\pi)^3 2E_5}.
\Ee

\section{results and discussion}

\begin{figure}[htbp]
\small
\centering
\includegraphics[scale=0.58]{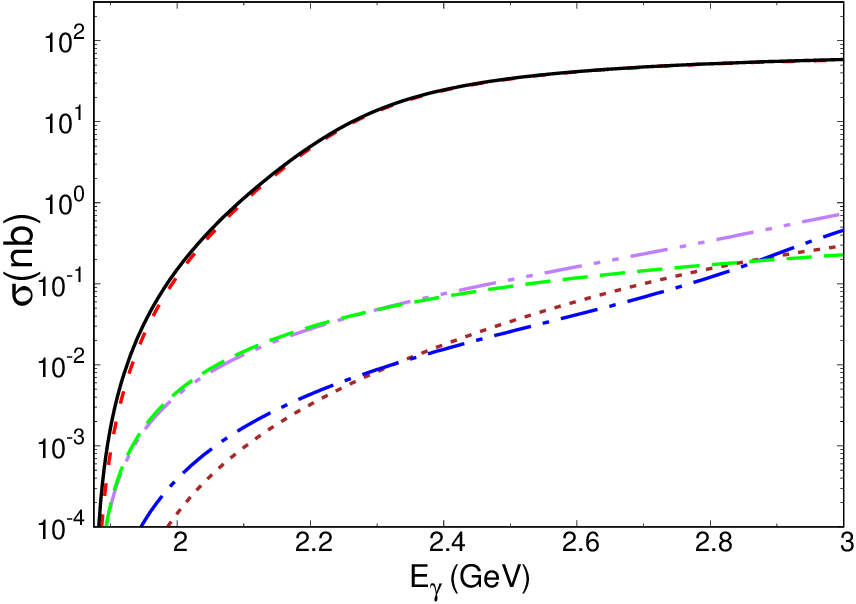}
\caption{Total cross section as a function of the beam energy $E_\gamma$ for the
${\gamma}p \to {\phi}{\pi}p$ reaction, without considering the $a_0$
exchange. The contributions from $\Delta(1232)$, $N(1535)$ and $N(1520)$, and $N(1440)$ are presented by the red-short-dashed,  purple-dash-dot-dotted,
blue-dash-dotted and brown-dotted lines, respectively.
The green-long-dashed line shows the $t$-channel background
contribution. Their sum is shown by the solid-black line.}
\label{xsection}
\end{figure}

Using the formalism from the preceding section, we have calculated the total and differential cross sections. We first consider the case without the $a_0$ exchange contribution. The results are presented in Fig.~\ref{xsection}. The figure details the contributions from various resonances and the $t$-channel background, showing that the $\Delta(1232)$ resonance production clearly dominates the reaction near the threshold. This dominance is primarily attributed to its large coupling to the $N\pi$ channel and its relatively low mass. In contrast, other nucleon resonances and the background term provide only minor contributions at these energies. However, with increasing energy, the roles of the $N(1535)$ and $N(1520)$ resonances become more significant. Interference effects among the individual amplitudes can be crucial for describing experimental data. In our model, most coupling constants are extracted from experimental decay widths, a method that determines their magnitudes but not their relative signs. Nevertheless, due to the dominance of the $\Delta(1232)$ in the energy region under study, we expect the interference effects to be insignificant. To verify this, we have altered the relative signs of the amplitudes by a factor of $-1$ and observed no significant changes in the results. It is also worth noting that our evaluation of the $\phi\rho\pi$ coupling constant uses the upper limit of the $\phi\to\rho\pi$ branching ratio suggested by the PDG~\cite{pdg2024}. This implies that the calculated background contribution is likely an upper estimate and could be even smaller in reality.

\begin{figure*}[htbp]
\small
\centering
\includegraphics[scale=0.8]{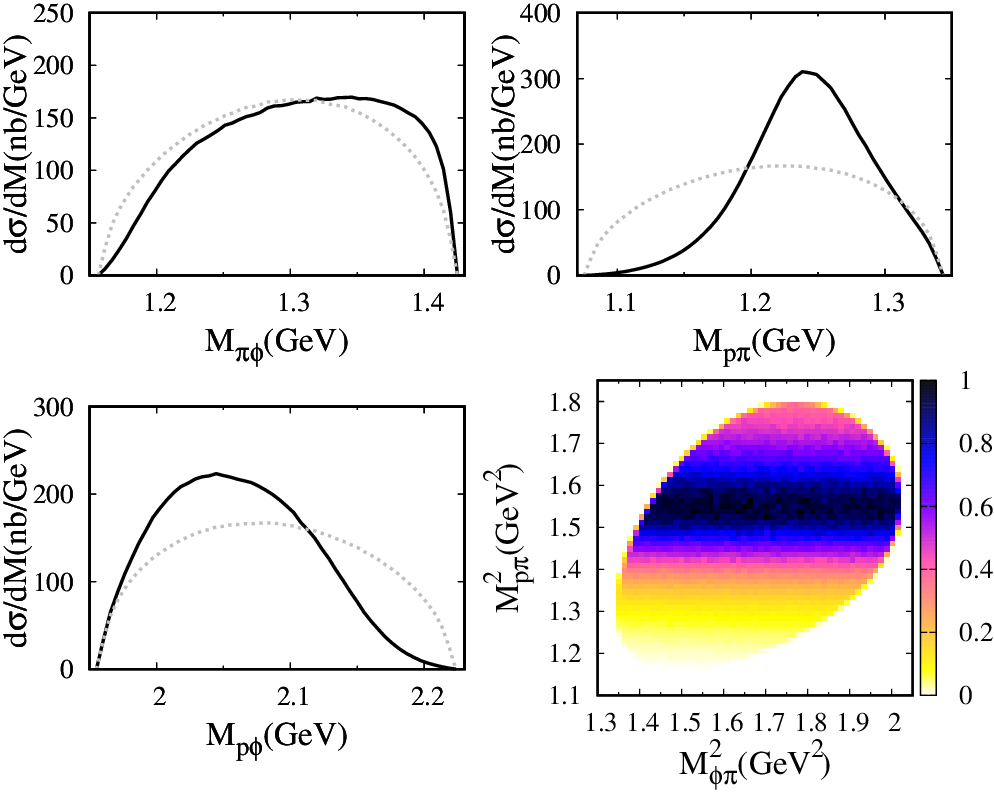}
\caption{Invariant mass spectra and Dalitz plot for the
${\gamma}p \to {\phi}{\pi}p$ reaction. The gray-dotted line in the invariant mass spectra show the corresponding phase space distributions for comparison.}
\label{IM}
\end{figure*}

In Fig.~\ref{IM}, we present our model's predictions for the invariant mass spectra of the $p\pi^0$, $\phi\pi^0$, and $p\phi$ systems, along with a Dalitz plot of $M^2_{p\pi^0}$ versus $M^2_{\phi\pi^0}$ for the $\gamma p \to \phi \pi^0 p$ reaction. In the spectra plots, the solid lines represent the model calculation, while the dotted lines show the phase space distributions for reference. As expected, the most prominent feature of the result is observed in the $p\pi^0$ invariant mass spectrum ($M_{p\pi^0}$, top-right panel). A clear and dominant peak is located at $M_{p\pi^0} \approx 1.23~\text{GeV}$, corresponding to the well-established $\Delta(1232)$ resonance. This indicates that the reaction proceeds overwhelmingly through the sequential decay channel $\gamma p \to \phi \Delta^+ \to \phi (p\pi^0)$. This conclusion is further  supported by the Dalitz plot analysis (bottom-right panel). The plot is dominated by a prominent horizontal band of high event density centered at $M^2_{p\pi^0} \approx 1.5$~GeV$^2$, which is caused by the $\Delta(1232)$ resonance. The less pronounced structures observed in the invariant mass spectra of the other particle pairs($\phi\pi$ and $p\phi$) are mainly due to kinematic reflections of the dominant $\Delta$ production process.

\begin{figure*}[htbp]
\small
\centering
\includegraphics[scale=0.8]{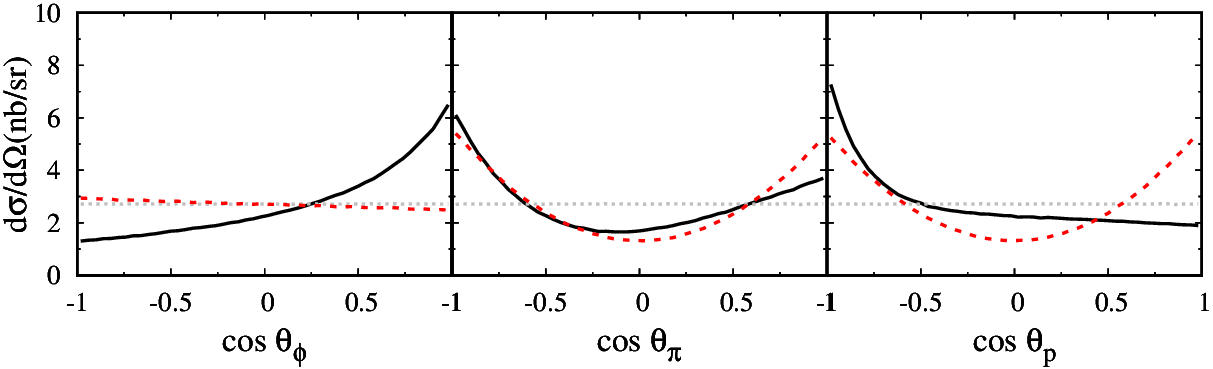}
\caption{Angular distributions of the final-state particles for the ${\gamma}p \to {\phi}{\pi}p$ reaction at $E_\gamma=2.5$ GeV. The black solid lines are obtained in the center-of-mass frame, taking the direction of the beam momentum as the z-axis. The red dashed line represents the angular distribution in the $\pi p$ rest frame using the Jackson frame. The gray dotted lines show the phase space distributions.}
\label{dsdx}
\end{figure*}

In addition to the invariant mass spectra and the Dalitz plot, we further investigate the reaction dynamics of $\gamma p \to \phi \pi^0 p$ by examining the angular distributions of the final-state particles, as shown in Fig.~\ref{dsdx}. In the calculations, the beam direction is taken as the $z$-axis.

The left panel of Fig.~\ref{dsdx} displays the differential cross section $d\sigma/d\Omega$ for the final $\phi$ as a function of $\cos\theta_\phi$. A prominent forward peak in the CM frame(solid line) is observed, indicating that the $\phi$ meson is preferentially produced in the direction of the incident photon beam. This forward-scattering behavior is a clear signal of a $t$-channel exchange process, which is consistent with the Feynman diagrams considered in our model. This feature can be verified by future experiments.

The middle and right panels of Fig.~\ref{dsdx} shows the angular distributions of the final $\pi^0$ and proton, respectively. The solid line, representing the results in the CM frame, shows a prominent backward enhancement. This feature can be attributed to a dominant $t$-channel meson exchange mechanism for the production of the $\Delta(1232)$ resonance, which subsequently decays and enhances the $\pi^0$ and $p$ yields at the backward angles. The dashed line represents the angular distributions in the $p\pi^0$ rest frame, calculated in the Jackson frame\cite{Jackson:1964zd}, where the z-axis is aligned with the initial proton's momentum. It exhibits a characteristic U-shaped, roughly symmetric distribution with a minimum near $\cos\theta_\pi \approx 0$ or $\cos\theta_p \approx 0$ and maxima in the forward and backward directions. This angular dependence is consistent with the decay of a spin-3/2 particle, i.e., the $\Delta(1232)$, into a spin-1/2 nucleon and a spin-0 $\pi$ meson.

\begin{figure*}[htbp]
\centering
\includegraphics[scale=0.9]{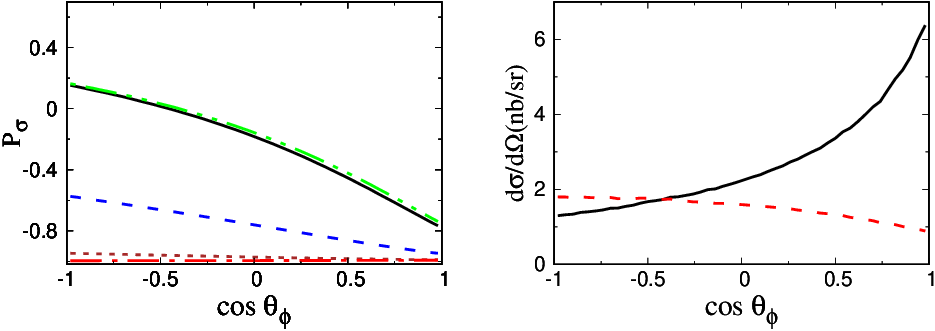}
\caption{Left: Predictions for the $P_{\sigma}$ asymmetry versus $\cos \theta_\phi$ in the center-of-mass frame, with the direction of the beam momentum as the z-axis. The lines correspond to different values of the coupling constant $g_{\Delta(1232)Na_0}$: $20.8$ (black solid line), $-20.8$ (green dash-dot-dotted line), $9.3$ (blue dashed line), $2.9$ (purple dotted line), and $0$ (red dash-dotted line). Right: Individual contributions to the $\phi$ angular distribution in the center-of-mass frame. The black solid line represents the $a_0$ exchange contribution (with $g_{\Delta(1232)Na_0}=20.8$), and the red dashed line represents the $\pi$ exchange contribution.}\label{psigma}
\end{figure*}

Next, we turn to the possible role of the $a_0$ exchange in this reaction. Evaluating this contribution requires two poorly constrained parameters: the $\Delta(1232)Na_0$ coupling constant, $g_{\Delta(1232)Na_0}$, and the cutoff parameter in the form factor. To date, both parameters are not well-established in the literature. It is possible to extract $g_{\Delta(1232)Na_0}$ from experimental data on the $\pi^+ p \to \Delta(1232)^{++} \eta$ reaction, where $t$-channel $a_0$ exchange is expected to contribute. However, this approach is unfeasible due to the significant uncertainties in the available data measured in the late 1970s\cite{Grether:1973sz} and the limited knowledge of competing $s$-channel processes, which involve the poorly known $\Delta^*\Delta\eta$ couplings. A reliable determination of $g_{\Delta(1232)Na_0}$ is therefore not possible through this channel at present. 

Given these challenges, we choose to estimate the impact of the $a_0$ exchange by relating its strength to the well-known $\pi^0$ exchange contribution in this reaction. Using a cutoff of $\Lambda_{a_0}=1.3$ GeV, we examine three scenarios by varying the coupling constant $g_{\Delta(1232)Na_0}$. Specifically, the values $g_{\Delta(1232)Na_0}=20.8$, $9.3$, and $2.9$ are chosen such that the $a_0$ exchange contribution is roughly 1/2, 1/10, and 1/100 of the $\pi^0$ exchange contribution, respectively. We have verified that even with the largest adopted value, $g_{\Delta(1232)Na_0}=20.8$, the resulting $a_0$ exchange contribution to the $\pi^+ p \to \eta \Delta^{++}$ reaction is compatible with current experimental data, given the aforementioned uncertainties.

With these assumptions about the magnitude of the $a_0$ exchange contribution, we can estimate its effects on various observables. To identify the role of the $a_0$ exchange in this reaction experimentally, it is essential to find an observable that can distinguish between different meson exchange contributions. According to the pioneering works in Refs. \cite{Schilling,Tannoudji}, the parity asymmetry was found to be suitable for distinguishing between natural and unnatural parity meson exchange contributions. In Ref.~\cite{Oh2006}, it was shown that this observable can be used to identify the scalar meson exchange contribution. The parity asymmetry is defined by the SDMEs as
\begin{equation}
P_\sigma=2\rho_{1 -1}^{1}-\rho_{0 0}^{1}.
\end{equation}
In Refs.~\cite{Schilling,Tannoudji}, it was shown that for the $\gamma N \to V N$ reaction, the parity asymmetry $P_\sigma$ equals 1 and -1 for natural and unnatural parity exchanges, respectively. Since the scalar meson $a_0$ has natural parity and the pseudoscalar meson $\pi$($\eta$) has unnatural parity, it is possible to distinguish their contributions by measuring this observable. It is worth noting that one difference between our work and the works in Ref.~\cite{Schilling,Tannoudji,Oh2006} is that our model includes baryon resonances in the intermediate state. However, the aforementioned features of $P_{\sigma}$ determined solely by the $\phi$-$\gamma$-Meson vertex when only scalar or pseudoscalar meson exchanges are involved. Therefore, one may expect these features to persist in our case. This argument can be verified by the numerical calculations.

In the left panel of Fig.~\ref{psigma}, we present the predictions for the $P_\sigma$ with $g_{\Delta(1232)Na_0}=20.8$ (black solid line), $9.3$ (blue dashed line), and $2.9$ (purple dotted line), respectively. For $g_{\Delta(1232)Na_0}=20.8$, the $P_\sigma$ is positive at backward angles, starting at about 0.1 at $\cos\theta_\phi = -1$. As $\cos\theta_\phi$ increasing, the $P_\sigma$ decreases gradually and reaches about -0.8 at $\cos\theta_\phi=1$. This feature can be understood by studying the angular distribution of $\phi$ in the center of mass frame caused by the $\pi$ and $a_0$ exchanges, which are shown in the right panel of Fig.~\ref{psigma}. As can be seen from the figure, the $\pi$ exchange (black solid line) causes a prominent forward enhancement, while the $a_0$ exchange (red dashed line) causes a slight backward enhancement. The backward enhancement for the $a_0$ exchange can be attributed to the relatively large mass of $a_0$ and the D-wave nature of the $\Delta(1232)Na_0$ coupling. This difference makes the $a_0$ exchange contribution more important at backward angles. Therefore, the $a_0$ exchange effect is best probed at backward angles in experiments.

If we take $g_{\Delta(1232)Na_0}=9.3$ (blue dashed line), the $a_0$ exchange contribution is significantly smaller than that of the $\pi$ exchange. In this case, $P_\sigma$ ranges from about $-0.5$ at the most backward angles to $-0.9$ at the most forward angles. When the coupling is further reduced to $g_{\Delta(1232)Na_0}=2.9$ (purple dotted line), $P_\sigma$ varies from approximately $-0.98$ to $-1$, closely matching the value of $-1$ (red dash-dotted line) expected for an almost pure unnatural-parity exchange (e.g., the dominant $\pi$ exchange). These findings highlight $P_\sigma$ as a valuable tool for identifying the $a_0$ exchange contribution in this reaction. It is important to note that the foregoing analysis does not incorporate uncertainties stemming from potential interference effects. To offer a preliminary assessment of these effects, we also computed the result with $g_{\Delta(1232)Na_0} =-20.8$ (green dash-dot-dotted line). This calculation indicates that interference effects on the overall result are relatively weak. Nevertheless, if the $\Delta(1232)Na_0$ coupling is very weak, for instance, on the order of $g_{\Delta(1232)Na_0} = 2.9$ or smaller, the uncertainties in the analysis may make it challenging to discern the $a_0$ exchange contribution and extract the coupling constant. In such cases, only an upper limit for this coupling might be estimable.

\section{summary}
In this work, we investigate the $\gamma p\to \phi\pi p$ reaction
using an effective Lagrangian approach. This analysis includes contributions from the $\Delta(1232)$, $N(1440)$, $N(1535)$ and
$N(1520)$ in the intermediate state and the background
term. Our results show that the production of the $\Delta(1232)$
dominates this reaction in the near threshold region. Particularly, we
examine the potential role of the $a_0(980)$ exchange in $\Delta(1232)$ excitation.
We demonstrate that the parity asymmetry $P_\sigma$ serves as an effective observable for identifying scalar exchange contributions and can be used to study the $\Delta(1232) N a_0$ coupling in this reaction.

\begin{acknowledgements}
We acknowledge the supports from the Natural Science Foundation of Shaanxi Province under Grants No.2024JC-YBMS-010.
\end{acknowledgements}

\end{document}